￥

\newcommand{\beq}{\begin{equation}}
\newcommand{\eeq}{\end{equation}}
\newcommand{\eq}[1]{\begin{align}#1\end{align}}
\newcommand{\lf}{\left}
\newcommand{\rf}{\right}
\newcommand{\nt}{\notag}

\documentclass[aps,prl,twocolumn,superscriptaddress]{revtex4-1}
\usepackage{graphicx}
\usepackage{verbatim}
\usepackage{mathrsfs}
\usepackage{cancel}
\usepackage{bm}
\usepackage{float}
\usepackage{color}
\usepackage{amsmath,amsfonts,amssymb}
\usepackage{hyperref}
\usepackage{cleveref}
\usepackage{epstopdf}

\begin{document}

\title{Electromagnetic effects induced by time-dependent axion field}

\author{Katsuhisa Taguchi}
\affiliation{Yukawa Institute for Theoretical Physics, Kyoto University, Kyoto 606-8502, Japan}

\author{Tatsushi Imaeda}
\affiliation{Department of Applied Physics, Nagoya University, Nagoya 464-8603, Japan}

\author{Tetsuya Hajiri}
\affiliation{Department of Materials Physics, Nagoya University, Nagoya 464-8603, Japan}

\author{Takuya Shiraishi}
\affiliation{Department of Physics, Nagoya University, Nagoya 464-8602, Japan}

\author{Yukio Tanaka}
\affiliation{Department of Applied Physics, Nagoya University, Nagoya 464-8603, Japan}

\author{Naoya Kitajima}
\affiliation{Department of Physics, Nagoya University, Nagoya 464-8602, Japan}

\author{Tatsuhiro Naka}
\affiliation{Department of Physics, Nagoya University, Nagoya 464-8602, Japan}
\affiliation{Kobayashi Masukawa Institute for the Origin of Particles and the Universe, Nagoya 464-8602, Japan}

\date{\today}
%

\begin{abstract}
We studied the dynamics of the so-called $\theta$-term, which exists in topological materials and is related to a hypothetical field predicted by Peccei-Quinn in particle physics, in a magnetic superlattice constructed using a topological insulator and two ferromagnetic insulators, 
where the ferromagnetic insulators had perpendicular magnetic anisotropies 
and different magnetic coercive fields.
We examined a way to drive the dynamics of the $\theta$-term in the magnetic 
superlattice through changing the inversion symmetry (from an anti-parallel to a parallel magnetic configuration) using an external magnetic field. 
As a result, we found that unconventional electromagnetic fields, which are magnetic field-induced charge currents and vice versa, are generated by the nonzero dynamics of the $\theta$-term. 
\end{abstract}

\maketitle

\textit{Introduction.}--
In the context of high-energy physics, an axion is an additional scalar degree of freedom, which gives a natural solution to the charge conjugation parity problem in the standard model of particle physics\cite{Peccei77}. In the cosmological context, it can also take the role of dark matter in the present universe, and the axion detection experiment is currently one of the most exciting fields of study. 
The axion couples with electromagnetic fields $\bm{E}$ and $\bm{B}$ as follows:
\begin{eqnarray} 
\label{eq:1} 
\mathcal{L}_{a} = \frac{\alpha}{2\pi } g \frac{a(t)}{f_a} \bm{E}\cdot\bm{B}, 
\end{eqnarray}
where $\alpha$ is a fine structure constant, 
$g$ is a coefficient of $\mathcal{O}(1)$ \cite{Dine81,Zhitnitsky80,J-E-Kim79,Shifman80}, $a(t)$ is the (time-dependent) axion field, and $f_a$ is the so-called Peccei-Quinn scale.
Because of an extremely small value of $a(t)/f_a \lesssim 10^{-19}$ in the present universe, it is difficult to detect the signature of a dark-matter axion. 

Intriguingly, a Lagrangian similar to Eq. (\ref{eq:1}) can be realized in topological materials: 
\begin{eqnarray} 
\mathcal{L}_{\theta} = \frac{e^2}{2\pi h} \theta \bm{E}\cdot\bm{B}, 
\end{eqnarray}
where $e$ is an elementary charge, and the term proportional to $\theta$ is 
the so-called $\theta$-term. This is analogous to $a(t)/f_a$ in Eq. (1), but $\theta$ is basically a static constant in the time-reversal symmetry.
The finite $\theta$ can be realized in the family of topological insulators (TIs), 
which includes magnetic-doped TIs\cite{Chang13,Hirahara17}, 
multilayers of magnetic TIs\cite{Mogi17,Q-L-He17}, 
Weyl semimetals (WSs)\cite{Huang15,S.Xu15,Lv15,Weng15}, 
and superlattices\cite{Burkov11,Tominaga14,Tominaga15}. 
In the presence of $\theta$, characteristic electromagnetic effects, \cite{Essin09,Qi08a,Qi09a,Karch09,Rosenberg10,Vazifeh10,Lan11,Tominaga15}, 
anomalous Hall effects\cite{R-Yu10,Nomura11,J-Wang15,Zyuzin12a,Okada16}, 
chiral magnetic effects\cite{Vilenkin80,Kharzeev08,Fukushima08,Vazifeh13,Sumiyoshi16,Sekine16,Taguchi16a,Li2016,Hayata17}, and Kerr effects\cite{Tse10,Maciejko10}, have been intensively studied.

One of the most interesting physical phenomena driven by the $\theta$-term is the electromagnetic effect via the dynamics of $\theta$ (i.e., $\partial_t \theta$), which could be analogous to the axion. So far, the unconventional optical effect \cite{Li10} and electric field-induced magnetic field \cite{Ooguri12} have been discussed under a nonzero $\partial_t \theta$, whose dynamics is caused by magnetic fluctuations, in materials with breaking time-reversal and inversion symmetries. Then, the time-average of $\partial_t \theta$ becomes zero, and its manipulation by an external field might be difficult.

\begin{figure}[ht]
\includegraphics[width=8.4cm]{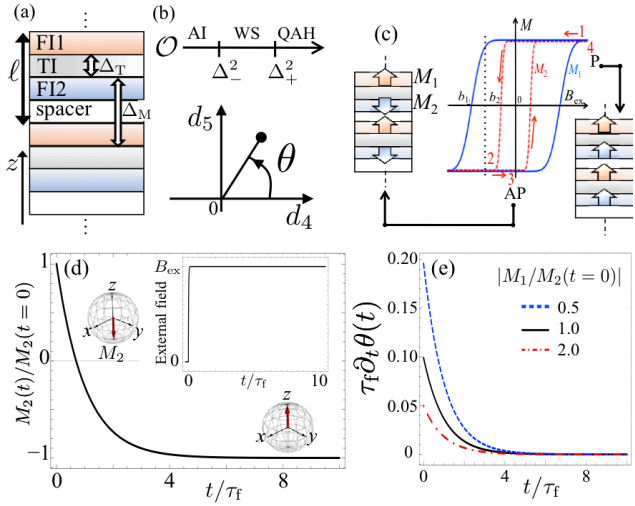}
\caption{ (Color online) 
(a) Superlattice constructed by [FI1/TI/FI2/spacer]$_n$.
(b) Phase diagram of superlattice, which is categorized by an order parameter $\mathcal{O}=M_1 M_2$ with several tunneling parameter 
$\Delta_{\pm} = \Delta_{\textrm{T}} \pm \Delta_{\textrm{M}}$. In the AI phase ($\mathcal{O} <\Delta_{-}^2$), 
the $\theta$-term is given by $\theta = \tan^{-1}{(d_5/d_4)}$ within the lowest order of $M_+$.
(c) Illustration of the magnetic hysteresis loops and magnetic configurations of the superlattice with different magnetic coercive fields  ($b_1$ and $b_2$) under an external magnetic field $\bm{B}_{\textrm{ex}}$ along the layered direction. $M_{1}$ and $M_2$ correspond to the magnetizations of the FIs. The antiparallel (AP) and parallel (P) magnetic configurations can be generated by $\bm{B}_{\textrm{ex}}$ though $1\to2\to3$ (red arrows) and $3\to4$, respectively. 
(d) Time-dependence of $M_2(t)$ by the external magnetic field $\bm{B}_{\textrm{ex}}$ (inset) during the $3\to 4$ process.
(e) Time-dependence of the dynamical $\theta$-term with several $|M_1/M_2(t=0)|$ at $M_1$ = 0.01 eV. 
}
\label{fig:1} 
\end{figure}

In this letter, we discuss a way to drive the dynamics of the $\theta$-term using an external magnetic field and consider the electromagnetic effects via $\partial_t \theta$ in a magnetic superlattice while breaking both the time- and inversion-reversal symmetries. The magnetic superlattice we consider is constructed using a TI and two ferromagnetic insulators (FIs) [FI1/TI/FI2/spacer]$_n$ [Fig. \ref{fig:1}(a)], where FI1 and FI2 have perpendicular magnetic anisotropies and different magnetic coercive fields. Here, to clearly define the $\theta$-term, we consider the axion insulator phase realized in the FI/TI superlattice [Fig. \ref{fig:1}(b)]. Then, $\theta$ can correspond to magnetic configurations of the FIs, where the inversion-reversal symmetry is preserved or broken corresponding to parallel (P) or anti-parallel (AP) magnetic configurations. These configurations could be controlled by an external magnetic field because of the different magnetic coercive fields. Through this control, the nonzero $\partial_t \theta$ is induced during the AP$\to$P process. Furthermore, we found unconventional electromagnetic effects under a nonzero $\partial_t \theta$. Unlike the conventional (static) electromagnetic effects, a dynamical magnetic field-induced charge current is generated and vice versa. 
%

\textit{Model.}--
We start with a model in the superlattice constructed by [FI1/TI/FI2/spacer]$_n$ as shown in Fig. \ref{fig:1}(a). 
Its effective Hamiltonian model can be described by \cite{Burkov11,Wang16a}
\begin{eqnarray}
&&H = \sum_{n,m}^{N} \int d\bm{k}_\parallel \psi^\dagger_{\bm{k}_\parallel,n} \mathcal{H}_{nm}(\bm{k}_\parallel) \psi_{\bm{k}_\parallel,m}, 
	\\
&&\mathcal{H}_{nm}(\bm{k}_\parallel)
	 = h_0 (\bm{k}_\parallel) \delta_{n,m} + h_1 \delta_{n,m+1} + h_2 \delta_{n,m-1},
	\\
&&h_0 (\bm{k}_\parallel)
	 = \hbar v \tau_z (\bm{\sigma} \times \bm{k}_\parallel)_z + \Delta_{\textrm{T}} \tau_x s_0
		+ \mathcal{M}, \\
&&h_1 = \frac{1}{2} \Delta_{\textrm{M}} (\tau_x + i\tau_y) s_0, 	
	\\
&&h_2 = \frac{1}{2} \Delta_{\textrm{M}} (\tau_x - i\tau_y) s_0, 	
	\\ 
&&\mathcal{M} = - \frac{1}{2} M_1 (\tau_0 + \tau_z) s_z - \frac{1}{2} M_2 (\tau_0 - \tau_z) s_z,
\end{eqnarray}
where the creation operator $\psi^\dagger_{\bm{k}_\parallel,n} = (\psi^\dagger_{+, \uparrow, \bm{k}_\parallel,n} \ \psi^\dagger_{+, \downarrow, \bm{k}_\parallel,n} \ \psi^\dagger_{-, \uparrow, \bm{k}_\parallel,n} \ \psi^\dagger_{-,\downarrow, \bm{k}_\parallel,n} )$ of an electron, where $\uparrow$ and $\downarrow$ are the spin indices and “+” and “--” indicate the top and bottom of the same TI layer, respectively. 
The labels $n$ and $m$ indicate the TI layers, and 
$\hbar \bm{k}_\parallel$ and $v$ are the momentum and velocity of the surface states of a TI layer, respectively. 
$\Delta_{\textrm{T}}$ ($\Delta_{\textrm{M}}$) denotes the tunneling between the top and bottom of the same TI layer (of neighboring TI layers). 
$s_0$ ($\tau_{0}$) and $s_{x,y,z}$ ($\tau_{x,y,z}$) are a $2\times2$ identical matrix and Pauli matrices acting on the spin (the surface) degrees of freedom, respectively. 
$\mathcal{M}$ indicates the exchange interaction of the FIs, where $M_1$ and $M_2$ correspond to the $z$-components of the magnetizations of FI1 and FI2, respectively. 
$\ell = L/N$ is the superlattice period, and $L$ is the length of the whole layer. 
In the limit $N\to \infty$, the effective Hamiltonian is also described by a Fourier transformed form as follows \cite{Burkov11,Wang16a}:
\begin{eqnarray}
\label{eq:3}
H = \int d\bm{k} \psi^\dagger_{\bm{k}} \lf[ \sum_{n=1}^5 d_{n}(\bm{k}) \Gamma^n + M_{+} \tau_0 s_z \rf] \psi_{\bm{k}} 
\end{eqnarray}
with 
$d_{n=1,2,3,4,5}=[ -\hbar v k_x, \hbar v k_y, - \Delta_{\textrm{M}} \sin{k_z \ell}, \Delta_{\textrm{T}} + \Delta_{\textrm{M}} \cos{k_z \ell}, M_{-}]$, 
, $\bm{k}=(k_x, k_y, k_z)$, $M_\pm \equiv (M_1\pm M_2)/2$, and $\Gamma^{n=1,2,3,4,5}=( \tau_z s_y, \tau_z s_x, \tau_y s_0, \tau_x s_0, \tau_z s_z)$, 
which is defined by $\{ \Gamma^n, \Gamma^m \} = 2\delta^{nm}$.

The dispersion of the Hamiltonian is given by 
$\epsilon^2 (\bm{k}) = \hbar^2 v^2 k_{\parallel}^2 + [ M_{+} \pm \sqrt{ M_{-}^2 + \Delta^2_{k_z} }]^2 $ with $\Delta^2_{k_z} = \Delta_{\textrm{M}}^2 + \Delta_{\textrm{T}}^2 + 2\Delta_{\textrm{M}} \Delta_{\textrm{T}} \cos{k_z \ell} $. 
Based on this, the gap-full state, $i.e.$, 
axion insulator (AI) and quantum anomalous Hall (QAH) systems are characterized by the conditions $\mathcal{O} > \Delta_-^2$ and $\mathcal{O} > \Delta_+^2$, respectively, where $\mathcal{O}\equiv M_{+}^2- M_{-}^2 = M_1 M_2$ is an 
order parameter in the superlattice [Fig. \ref{fig:1}(b)]. 
The gapless phase (WS phase) is realized 
for $\Delta_-^2 < \mathcal{O} < \Delta_+^2$.

The $\theta$-term was studied in the AI phase and WS phase. The former is clearly described by $\delta\theta = \tan^{-1}(d_5/d_4)$ within the lowest order of $M_+$ [Fig. \ref{fig:1}(b)], and the latter is given by the distance 
of each point-node of the dispersion\cite{Burkov11}. 
In the following, to discuss the dynamical $\theta$-term clearly, 
we focus on the AI phase.

\textit{Electromagnetic effect via $\delta\theta$.}--
The electromagnetic effects in the AI phase can be given by\cite{Sikivie83,Sikivie85,Essin09,Qi08a,Qi09a,Karch09} 
\begin{eqnarray} \nt
\mathcal{L} && = \mathcal{L}_{\textrm{Maxwell}} + \mathcal{L}_{\theta} + \mathcal{L}_{\textrm{e}}
	\\ \label{eq:Lagrangian}
	&& = \frac{1}{2} \bigl[ \epsilon E^2 - \frac{1}{\mu} B^2\bigr] 
		+\frac{e^2}{2\pi h} (\theta_0 + \delta \theta) \bm{E}\cdot\bm{B}
		+ \bm{j}_e \cdot \bm{A}.
\end{eqnarray}
Here, $ \mathcal{L}_{\textrm{e}}$ denotes the gauge coupling between charge current $\bm{j}_e$ and vector potential $\bm{A}$. $\epsilon$ and $\mu$ are 
the electrical permittivity and magnetic permeability of the medium, respectively. 
$\theta_0$ ($= 0$ or $\pi$) is a static value 
and $\delta\theta$ denotes the deviation from $\theta_0$ \cite{Li10,Wang11,Ooguri12,Lee15,Wang16a}. 
Below, we simply consider $\delta\theta$ in the linear response of $d_5$ and $d_4 \sim \Delta_+$. 
Then, $\delta\theta$ is given by \cite{Sekine16,Wang16a}
\begin{eqnarray} 
\label{eq:theta-1}
\delta \theta = \tan^{-1}{\lf[ \frac{M_-}{\Delta_+} \rf] }.
\end{eqnarray}
The Maxwell equations are given by 
\eq{\begin{split}
&\bm{\nabla} \times \bm{E} = - \partial_t \bm{B} 
	\\ \label{eq: Maxwell}
&\bm{\nabla} \times \bm{B}/\mu = \epsilon \partial_t \bm{E} + \bm{j}_e + \bm{j}_\theta, 
\end{split}
}
where $\bm{j}_\theta = - [ (\bm{\nabla} \delta \theta ) \times \bm{E} + (\partial_t \delta \theta) \bm{B} ] c \epsilon \alpha/\pi $ is a charge current due to $\delta \theta$, and $c$ is the velocity of light. 
Then, the wave equation becomes 
\eq{
(\partial_t^2 - c^2 \nabla^2 ) \bm{B} & = \frac{1}{\epsilon}\bm{\nabla}\times (\bm{j}_e + \bm{j}_\theta).
}
The right-hand side of the above equation indicates the source term for the wave equation. The first term is conventional but the second term, which depends on $\bm{\nabla} \delta \theta$ and $ \partial_t \delta \theta$, is unconventional and
is given by 
$\bm{\nabla}\times \bm{j}_\theta/\epsilon = 
-\frac{c \alpha}{\pi} \lf[ (\bm{\nabla}\cdot \bm{E}) \bm{\nabla}\delta\theta - (\bm{\nabla}\delta\theta)\cdot \bm{\nabla} \bm{E} \rf]
 -\frac{c \alpha \mu}{\pi} \partial_t \delta \theta \lf(\epsilon \partial_t \bm{E} + \bm{j}_e + \bm{j}_\theta \rf).$
In particular, for $\bm{\nabla}\delta\theta=0$ and $\bm{\nabla}\times \bm{j}_e=0$, the wave equation becomes 
\eq{ \label{eq:wave equation 2}
(\partial_t^2 - c^2 \nabla^2 ) \bm{B} & = -\frac{c \alpha\mu }{\pi} \partial_t \delta \theta \lf( \epsilon \partial_t \bm{E} + \bm{j}_e + \bm{j}_\theta \rf).
}

\textit{Dynamics of $\delta\theta$.}--
To drive the nonzero dynamics of $\theta$ ($\partial_t \delta \theta$), we focus on magnetic configurations in the superlattice. 
Here, we assume that the magnetizations of FI1 and FI2 have perpendicular magnetic anisotropies and different magnetic coercive fields $b_1$ and $b_2$, as illustrated in Fig. \ref{fig:1}(c). 
Then, an external magnetic field $\bm{B}_{\textrm{ex}} =B_{\textrm{ex}} \bm{\hat{z}}$ along the layered direction can trigger a change in the magnetic configurations ($\bm{\hat{z}}$ is a unit vector).
For example, first we set the AP magnetic configuration, which can be generated through the $1\to2\to3$ process, with $ b_1 <B_{\textrm{ex}} < b_2$, as shown in Fig. \ref{fig:1}(c). 
Then, the AP magnetic configuration becomes stable even when $B_{\textrm{ex}}=0$.
After $B_{\textrm{ex}}\neq 0$ is applied, the $M_2$ values of the FIs are changed by $B_{\textrm{ex}}$, and the magnetic configuration can be dynamically changed from the AP configuration to the P magnetic configuration via the $3\to4$ process.
The changes in the magnetic configurations are caused by the change in the magnetization of FI2, which is switched from the $-z$ to the $+z$ direction during $3\to 4$. 
The above process could drive the nonzero $\partial_t \delta \theta$.

The magnetization switch can be phenomenologically understood using the Bloch equation for the magnetization motion as $\partial_t M_{2} (t) = [M_{2}(t=0) - M_{2}(t)]/\tau_{\textrm{f}}$, where $\tau_{\textrm{f}}$ is the characteristic relaxation time of the magnetization switching, which depends on the material. To solve this equation, we assume the initial condition $M_2(t=0) = -M_2$ with $M_2>0$, as illustrated in Fig. \ref{fig:1}(d). 
Then, we have $M_2(t) = M_2 [1-2 \exp{(-t/\tau_{\textrm{f}}) }] $ for $M_- (t) = [ M_1 - M_2(t)]/2 $ [Fig. \ref{fig:1}(d)]. As a result, $\partial_t \delta \theta$ is given by Eq. (\ref{eq:theta-1}) as follows:
\eq{\label{eq:time-dependent axion field}
\partial_t \delta\theta (t) & = \frac{ \Delta_+ M_2 \exp{(-t/\tau_{\textrm{f}}) } }{ \Delta_+^2 + M_-^2 (t) } \frac{1}{\tau_\textrm{f}}. 
}
Using the $\partial_t \delta \theta$ [Fig. \ref{fig:1}(e)], we consider the following three cases. Below, we simply focus on the time-dependence of $\bm{B}$ and $\partial_t\delta \theta$, and we ignore $c^2 \nabla^2 \bm{B}$ and set $\bm{\nabla}\delta\theta=0$. 
\begin{figure}[b]
\includegraphics[width=8.5cm]{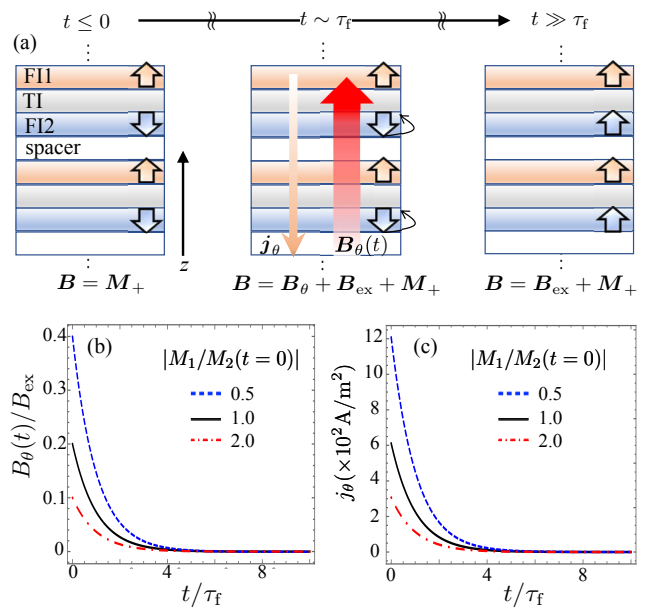}
\caption{ (Color online) 
Dynamical $\theta$ term induces magnetic field $\bm{B}_\theta$ and charge current $\bm{j}_\theta$.
(a) Illustration of $\bm{B}_\theta$ (red arrows) and $\bm{j}_\theta$ (orange arrows) in the presence of the dynamical $\theta$ term during the $3(t\leq 0) \to 4 (t\gg t_{\textrm{f}})$ process.
$\bm{B}_{\textrm{ex}}$ is an external magnetic field and $\bm{M}_+ = [M_1 + M_2 (t)] \hat{\bm{z}}$ corresponds to the total magnetization of the FIs.
(b) Time-dependence of $B_\theta$ and (c) time-dependence of $j_\theta$. 
We use the parameters $\Delta_+ = 0.1$ eV, $M_1=10$ meV, $\tau_\textrm{f}=1$ ns, $\epsilon=8.8\times10^{-12}$F/m, and $B_{\textrm{ex}}=0.1$ T.
}
\label{fig:2} 
\end{figure}

\textit{ $\partial_t \delta \theta$-induced magnetic field and charge current.}-- 
We consider the responses by the dynamical $\theta$ term $\partial_t \delta\theta (t)$ during the $3\to4$ process under a static external magnetic field $\bm{B}_{\textrm{ex}} = B_{\textrm{ex}} \hat{\bm{z}}$, which is applied at $t=0$.
Here, we focus on the unconventional magnetic field [$\bm{B}_\theta(t)$] via a nonzero $\partial_t \delta \theta$, which can be defined and decomposed as $\bm{B}(t) \equiv \bm{B}_{\textrm{ex}} + \bm{M}_+(t) + \bm{B}_\theta(t)$.
Then, the time-revolution of $\bm{B}_\theta(t)$ is given from $\partial_t^2 \bm{B}(t) = \lf[ \frac{\alpha}{\pi} \partial_t \delta \theta \rf]^2 \bm{B} (t)$: 
\eq{ \label{eq:wave equation caseA}
\partial_t^2 \bm{B}_\theta
	& = \frac{\alpha^2}{\pi^2} (\partial_t \delta \theta)^2(\bm{B}_\theta +\bm{B}_{\textrm{ex}} + \bm{M}_+) - \partial_t^2 \bm{M}_+.
}
To solve this, we use the following final and initial conditions:
\eq{\begin{split}
&\bm{B}_\theta ( t \to +\infty ) = 0
	\\ \label{eq: Initial condition}
&\partial_t \bm{B}_\theta (t=0) = - \partial_t \bm{M}_+(t)|_{t\to0}, 
\end{split}
}
where the former indicates the removal of a divergent solution, and the latter is determined from $\partial_t \bm{B}(t) = - \bm{\nabla}\times \bm{E}$ at $t=0$, while assuming $\bm{\nabla}\times \bm{E} \to 0$ at $t=0$.

Figure \ref{fig:2} illustrates the dynamical $\theta$-term 
that induces magnetic field $\bm{B}_\theta$. 
The nonzero $\bm{B}_\theta$ is only generated during the $3\to 4$ process. 
Furthermore, from Eq. (\ref{eq: Maxwell}), the $\partial_t \delta \theta$ couples with the external magnetic field and drives the charge current $\bm{j}_{\theta} = - [\partial_t (\delta \theta)] \bm{B} c \epsilon \alpha/\pi $ only during $3\to4$, as described in Fig. \ref{fig:2}.
This magnetic-field induced charge current can be regarded as a kind of chiral magnetic effect\cite{Vilenkin80,Kharzeev08,Fukushima08,Vazifeh13,Sumiyoshi16,Sekine16,Taguchi16a,Li2016,Hayata17}.

\begin{figure}[t]
\includegraphics[width=8.7cm]{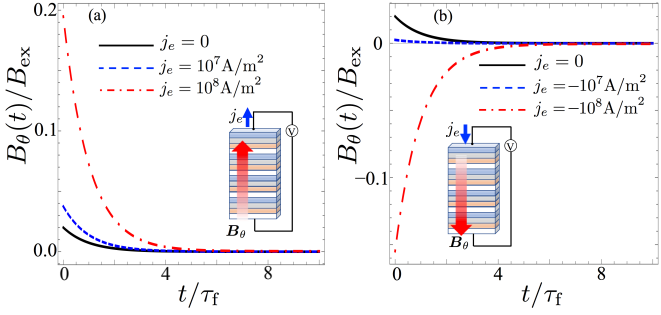}
\caption{ (Color online) 
Time-dependence of magnetic field $B_{\theta} \equiv B_{\theta}(j_e=0) + B_{\theta}(j_e\neq 0)$ due to $\partial_t \delta \theta$ in presence of external magnetic field $\bm{B}_{\textrm{ex}}$ and charge current $\bm{j}_e=j_e \hat{\bm{z}}$ along layered direction in (a) $j_e>0$ and (b) $j_e<0$, with several $j_e$ at $|M_1/M_2(t=0)|=1$. The parameters are the same as those used in Fig. \ref{fig:2}.
}
\label{fig:3} 
\end{figure}

\textit{Under charge current along $\bm{B}_{\textrm{ex}}$.}--
Next, we discuss the magnetic field via $\partial_t \delta \theta$ in the presence of both the magnetic field $\bm{B}_{\textrm{ex}}= B_{\textrm{ex}}\bm{\hat{z}}$ and the charge current $\bm{j}_e=j_e \bm{\hat{z}}$, which is spatially uniform.
Then, the time-revolution of $\bm{B}_\theta = B_\theta \hat{\bm{z}} $ becomes 
\eq{\nt
\partial_t^2 B_\theta
	& = \frac{\alpha^2}{\pi^2} (\partial_t \delta \theta)^2(B_{\theta} + B_{\textrm{ex}} + M_+) - \partial_t^2 M_+ 
		\\ \label{eq: wave equation 2}
		& -\frac{c \alpha\mu_0}{\pi} \partial_t \delta \theta j_e.
}
$B_\theta \equiv B_\theta(j_e=0) + B_\theta(j_e\neq0) $ is generated not only by the coupling between $\partial_t \delta \theta$ and $\bm{B}_{\textrm{ex}}$ but also by the coupling between $\partial_t \delta \theta$ and $\bm{j}_e$. The former $B_\theta(j_e=0)$ is given in Fig. \ref{fig:2}. 
The latter $B_\theta(j_e\neq0)$ will be dominant when the magnitude of $j_e$ is sufficiently large (Fig. \ref{fig:3}).
It should be noted that during the $3\to4$ process, the charge current via the chiral magnetic effect $j_\theta$ is also generated, and the charge current $j_e$ includes $j_\theta$ as $j_e \to j_e + j_\theta$.
However, the magnitude of $j_\theta$ is smaller than the sufficient value of current (e.g., $10^7$A/m$^2$) [see Fig. \ref{fig:2}(c)]. Hence, the small $j_\theta$ hardly affects $B_\theta$.

\begin{figure}[b]
\includegraphics[width=8.8cm]{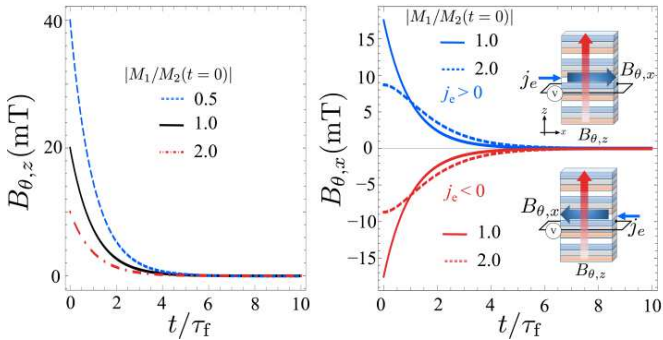}
\caption{ (Color online) 
Magnetic field $\bm{B}_{\theta} = B_{\theta,z}\hat{\bm{z}} + B_{\theta,x}\hat{\bm{x}} $ induced by dynamics of $\theta$ under charge current along $x$ direction ($\bm{j}_e = j_e \hat{\bm{x}}$). (a) Time-dependence of $B_{\theta,z}$ with several $|M_1/M_2|$ [see also Fig.\ref{fig:2}(b)]. (b) Time-dependence of $B_{\theta,x}$ at $j_e = \pm 10^7$A/m$^2$ with $|M_1/M_2(t=0)|=1, 2$. The parameters are the same as those used in Fig. \ref{fig:2}. (inset) Illustration of the magnetic fields due to the dynamics of $\theta$.
}
\label{fig:4} 
\end{figure} 

\textit{Charge current perpendicular to $\bm{B}_{\textrm{ex}}$.}--
We consider $\bm{B}_\theta$ in the presence of the external magnetic field $\bm{B}_{\textrm{ex}}=B_{\textrm{ex}} \hat{\bm{z}}$ and the charge current $\bm{j}_e=j_e \bm{\hat{z}}$.
Then, we have 
\eq{ \label{eq:wave equation caseD1}
\partial_t^2 B_{\theta,z} 
	& = \frac{\alpha^2}{\pi^2} (\partial_t \delta \theta)^2( B_{\theta,z} + B_{\textrm{ex}} + M_+) - \partial_t^2 M_+, 
	\\	\label{eq:wave equation caseD2}
\partial_t^2 B_{\theta,x} 
	& =-\frac{c \alpha\mu_0}{\pi} \partial_t \delta \theta j_e.
}
The induced magnetic field $\bm{B}_\theta= B_{\theta,z}\bm{\hat{z}} + B_{\theta,x}\bm{\hat{x}}$ is solved under the initial condition [Eq. (\ref{eq: Initial condition})], and it can be decomposed into two terms. 
$B_{\theta,z}$ is the same magnetic field $B_\theta(j_e=0)$ parallel to $\bm{B}_{\textrm{ex}}$ [Fig. \ref{fig:4}(a)], and $B_{\theta,x}$ corresponds to the same magnetic field $B_\theta(j_e\neq0)$ parallel to $\bm{j}_{e}$ [Fig. \ref{fig:4}(b)].

It should be noted that $B_{\theta,x}$ is the current-induced magnetic field. The direction of $B_{\theta,x}$ is parallel to the applied charge current direction, which is different from that of the
charge currents via Amp\'ere's law. In addition, $B_{\theta,x}$ is regarded as a kind of electromagnetic effect (current-induced magnetic field) or Edelstein effect.
The conventional current-induced magnetic field\cite{DzyaloshinskiiI60,Astrov60,Fiebig05} is represented by $\mathcal{B}_{m}=\chi_{mn} j_{e,n}$, where $\mathcal{B}_{m}$ is an induced magnetic field, $j_{e,n}$ is a static charge current, and $\chi_{mn}$($m,n=x,y,z$) is a coefficient that is usually determined by the characteristic parameters in an equilibrium state in materials. Hence, $\chi_{mn}$ is static.
On the other hand, $B_{\theta,x}(t)$ is described by the form $B_{\theta,x}=\chi_{\theta,xn}(t) j_{e,n}$, where the coefficient $\chi_{\theta,xn}$ is represented by 
\eq{
\chi_{\theta,xn} (t)& = -\frac{c \alpha\mu_0}{\pi}\delta_{xn} \biggl[ \int_{0}^t dt_1 \delta \theta(t_1) +c_1 t + c_2 \biggr]. 
}
Here, $c_1$ and $c_2$ are arbitrary constants determined by the initial condition. 
Because the theta-term $\delta \theta$ is dynamical, $\chi_{\theta,xn}$ is dynamical, unlike the conventional $\chi_{mn}$.

Finally, we discuss an experimental instrument for the obtained results in the response to the dynamics of $\theta$. 
Although we consider the model in the superlattice [FI1/TI/FI2/spacer]$_n$ to study a simple model of $\partial_t\theta$, 
the number of ferromagnetic metals is practically larger than that of the FIs. 
Recently, for example, an axion-insulator multilayer constructed of a TI film and Cr-doped TI film with different magnetic coercive fields was experimentally studied\cite{Mogi17}.
A superlattice of tailored axion-insulators could be used for the model we considered.
In such materials, it is expected that the manipulation of the magnetic configurations of FIs (AP and P) is driven by an external magnetic field along the layered direction, as illustrated in Fig. \ref{fig:2}(a). As a result, the 
dynamics of $\theta$ can be driven via the AP $\to$ P process.
Then, three electromagnetic effects occur during the process.
First, the chiral magnetic effect, which is a magnetic field-induced charge current, could be expected. 
In other words, in the absence of an applied charge current, the chiral magnetic effect, which is caused by the dynamical $\theta$-term and applied magnetic field, drives a charge current of $\approx 10^2$ A/m$^2$ [Fig. \ref{fig:2}(c)], which could be detectable. 
Second, another electromagnetic effect is $B_{\theta,z}$, which is driven by the dynamics of the $\theta$-term and the external magnetic field $\bm{B}_{\textrm{ex}}$. The magnitude of $B_{\theta,z}$, for example, becomes $\approx 20$ mT during a few tens of $\tau_\textrm{f}$ [Fig. \ref{fig:4}(a)].
We expect that it could be difficult to distinguish $B_{\theta,z}$ from the total magnetic field $\bm{B}$ [Fig. \ref{fig:2}(a)] because the magnetic fields are in the same direction.
Third, the obtained electromagnetic effect is the current-induced magnetic field $B_{\theta,x}$, which occurs when we apply the charge current.
Under the in-plane charge current $\bm{j}_e$, $B_{\theta,x}$, which is along the applied current direction, reaches $\approx \pm 10 $mT when $M_1=M_2=10$ meV and $d_4=0.1$ eV with $j_e = \pm 10^{7}$ A/m$^2$. 
This current-induced magnetic field could be measurable by manipulations of the current direction.
It should be noted that under the in-plane charge current, a spin-transfer-torque-induced magnetic field\cite{Garate10,Yokoyama10a,Sakai14} could also be generated by the coupling between the applied current and magnetizations on the surfaces of the TIs. However, this magnetic field could be much smaller than that of $B_{\theta,x}$, which would make it experimentally negligible.

\textit{Conclusion.}--
We have theoretically studied a way to drive the dynamics of $\theta$ in a magnetic superlattice with an axion insulator phase. 
As a result, we have shown the unconventional electromagnetic effects via $\partial_t \theta$: 
The dynamical magnetic field-induced charge current, which is a kind of chiral magnetic effect, and the current-induced magnetic effect, which is a kind of Edelstein effect, have been discussed. 
The obtained results, including the artificial control of $\partial_t \theta$ by an external field, could be analogous to the time-dependent axion field $a(t)/f_a$ in dark-matter axion physics.

\begin{acknowledgments}
This work was supported by the institute for advanced research (IAR) in Nagoya University through a Grant-in-Aid. 
In addition, this work was supported
by a Grant-in-Aid for Scientific Research on Innovative
Areas, Topological Material Science (No. JP15H05853), 
and a Grant-in-Aid for Challenging Exploratory 
Research (Grant No. JP15K13498 and 16K13803) 
from the Ministry of Education, Culture, Sports, Science, and
Technology, Japan (MEXT). 
K.T. also gives thanks for the support of the Core Research for Evolutional Science and Technology (CREST) of the Japan Science and Technology Corporation (JST) [JPMJCR14F1]. 
N.K. acknowledges the support by Grant-in-Aid for JSPS Fellows.

\end{acknowledgments}

\bibliography{Axion_v1}

\begin{thebibliography}{52}%
\makeatletter
\providecommand \@ifxundefined [1]{%
 \@ifx{#1\undefined}
}%
\providecommand \@ifnum [1]{%
 \ifnum #1\expandafter \@firstoftwo
 \else \expandafter \@secondoftwo
 \fi
}%
\providecommand \@ifx [1]{%
 \ifx #1\expandafter \@firstoftwo
 \else \expandafter \@secondoftwo
 \fi
}%
\providecommand \natexlab [1]{#1}%
\providecommand \enquote  [1]{``#1''}%
\providecommand \bibnamefont  [1]{#1}%
\providecommand \bibfnamefont [1]{#1}%
\providecommand \citenamefont [1]{#1}%
\providecommand \href@noop [0]{\@secondoftwo}%
\providecommand \href [0]{\begingroup \@sanitize@url \@href}%
\providecommand \@href[1]{\@@startlink{#1}\@@href}%
\providecommand \@@href[1]{\endgroup#1\@@endlink}%
\providecommand \@sanitize@url [0]{\catcode `\\12\catcode `\$12\catcode
  `\&12\catcode `\#12\catcode `\^12\catcode `\_12\catcode `\%12\relax}%
\providecommand \@@startlink[1]{}%
\providecommand \@@endlink[0]{}%
\providecommand \url  [0]{\begingroup\@sanitize@url \@url }%
\providecommand \@url [1]{\endgroup\@href {#1}{\urlprefix }}%
\providecommand \urlprefix  [0]{URL }%
\providecommand \Eprint [0]{\href }%
\providecommand \doibase [0]{http://dx.doi.org/}%
\providecommand \selectlanguage [0]{\@gobble}%
\providecommand \bibinfo  [0]{\@secondoftwo}%
\providecommand \bibfield  [0]{\@secondoftwo}%
\providecommand \translation [1]{[#1]}%
\providecommand \BibitemOpen [0]{}%
\providecommand \bibitemStop [0]{}%
\providecommand \bibitemNoStop [0]{.\EOS\space}%
\providecommand \EOS [0]{\spacefactor3000\relax}%
\providecommand \BibitemShut  [1]{\csname bibitem#1\endcsname}%
\let\auto@bib@innerbib\@empty
\bibitem [{\citenamefont {Peccei}\ and\ \citenamefont
  {Quinn}(1977)}]{Peccei77}%
  \BibitemOpen
  \bibfield  {author} {\bibinfo {author} {\bibfnamefont {R.~D.}\ \bibnamefont
  {Peccei}}\ and\ \bibinfo {author} {\bibfnamefont {H.~R.}\ \bibnamefont
  {Quinn}},\ }\href {\doibase 10.1103/PhysRevLett.38.1440} {\bibfield
  {journal} {\bibinfo  {journal} {Phys. Rev. Lett.}\ }\textbf {\bibinfo
  {volume} {38}},\ \bibinfo {pages} {1440} (\bibinfo {year}
  {1977})}\BibitemShut {NoStop}%
\bibitem [{\citenamefont {Dine}\ \emph {et~al.}(1981)\citenamefont {Dine},
  \citenamefont {Fischler},\ and\ \citenamefont {Srednicki}}]{Dine81}%
  \BibitemOpen
  \bibfield  {author} {\bibinfo {author} {\bibfnamefont {M.}~\bibnamefont
  {Dine}}, \bibinfo {author} {\bibfnamefont {W.}~\bibnamefont {Fischler}}, \
  and\ \bibinfo {author} {\bibfnamefont {M.}~\bibnamefont {Srednicki}},\
  }\href@noop {} {\bibfield  {journal} {\bibinfo  {journal} {Physics Letters
  B}\ }\textbf {\bibinfo {volume} {104}},\ \bibinfo {pages} {199} (\bibinfo
  {year} {1981})}\BibitemShut {NoStop}%
\bibitem [{\citenamefont {Zhitnitsky}(1980)}]{Zhitnitsky80}%
  \BibitemOpen
  \bibfield  {author} {\bibinfo {author} {\bibfnamefont {A.~R.}\ \bibnamefont
  {Zhitnitsky}},\ }\href@noop {} {\bibfield  {journal} {\bibinfo  {journal}
  {Sov. J. Nucl. Phys.}\ }\textbf {\bibinfo {volume} {31}},\ \bibinfo {pages}
  {260} (\bibinfo {year} {1980})}\BibitemShut {NoStop}%
\bibitem [{\citenamefont {Kim}(1979)}]{J-E-Kim79}%
  \BibitemOpen
  \bibfield  {author} {\bibinfo {author} {\bibfnamefont {J.~E.}\ \bibnamefont
  {Kim}},\ }\href {\doibase 10.1103/PhysRevLett.43.103} {\bibfield  {journal}
  {\bibinfo  {journal} {Phys. Rev. Lett.}\ }\textbf {\bibinfo {volume} {43}},\
  \bibinfo {pages} {103} (\bibinfo {year} {1979})}\BibitemShut {NoStop}%
\bibitem [{\citenamefont {Shifman}\ \emph {et~al.}(1980)\citenamefont
  {Shifman}, \citenamefont {Vainshtein},\ and\ \citenamefont
  {Zakharov}}]{Shifman80}%
  \BibitemOpen
  \bibfield  {author} {\bibinfo {author} {\bibfnamefont {M.}~\bibnamefont
  {Shifman}}, \bibinfo {author} {\bibfnamefont {A.}~\bibnamefont {Vainshtein}},
  \ and\ \bibinfo {author} {\bibfnamefont {V.}~\bibnamefont {Zakharov}},\
  }\href {\doibase 10.1016/0550-3213(80)90209-6} {\bibfield  {journal}
  {\bibinfo  {journal} {Nuclear Physics B}\ }\textbf {\bibinfo {volume}
  {166}},\ \bibinfo {pages} {493} (\bibinfo {year} {1980})}\BibitemShut
  {NoStop}%
\bibitem [{\citenamefont {Chang}\ \emph {et~al.}(2013)\citenamefont {Chang},
  \citenamefont {Zhang}, \citenamefont {Feng}, \citenamefont {Shen},
  \citenamefont {Zhang}, \citenamefont {Guo}, \citenamefont {Li}, \citenamefont
  {Ou}, \citenamefont {Wei}, \citenamefont {Wang}, \citenamefont {Ji},
  \citenamefont {Feng}, \citenamefont {Ji}, \citenamefont {Chen}, \citenamefont
  {Jia}, \citenamefont {Dai}, \citenamefont {Fang}, \citenamefont {Zhang},
  \citenamefont {He}, \citenamefont {Wang}, \citenamefont {Lu}, \citenamefont
  {Ma},\ and\ \citenamefont {Xue}}]{Chang13}%
  \BibitemOpen
  \bibfield  {author} {\bibinfo {author} {\bibfnamefont {C.-Z.}\ \bibnamefont
  {Chang}}, \bibinfo {author} {\bibfnamefont {J.}~\bibnamefont {Zhang}},
  \bibinfo {author} {\bibfnamefont {X.}~\bibnamefont {Feng}}, \bibinfo {author}
  {\bibfnamefont {J.}~\bibnamefont {Shen}}, \bibinfo {author} {\bibfnamefont
  {Z.}~\bibnamefont {Zhang}}, \bibinfo {author} {\bibfnamefont
  {M.}~\bibnamefont {Guo}}, \bibinfo {author} {\bibfnamefont {K.}~\bibnamefont
  {Li}}, \bibinfo {author} {\bibfnamefont {Y.}~\bibnamefont {Ou}}, \bibinfo
  {author} {\bibfnamefont {P.}~\bibnamefont {Wei}}, \bibinfo {author}
  {\bibfnamefont {L.-L.}\ \bibnamefont {Wang}}, \bibinfo {author}
  {\bibfnamefont {Z.-Q.}\ \bibnamefont {Ji}}, \bibinfo {author} {\bibfnamefont
  {Y.}~\bibnamefont {Feng}}, \bibinfo {author} {\bibfnamefont {S.}~\bibnamefont
  {Ji}}, \bibinfo {author} {\bibfnamefont {X.}~\bibnamefont {Chen}}, \bibinfo
  {author} {\bibfnamefont {J.}~\bibnamefont {Jia}}, \bibinfo {author}
  {\bibfnamefont {X.}~\bibnamefont {Dai}}, \bibinfo {author} {\bibfnamefont
  {Z.}~\bibnamefont {Fang}}, \bibinfo {author} {\bibfnamefont {S.-C.}\
  \bibnamefont {Zhang}}, \bibinfo {author} {\bibfnamefont {K.}~\bibnamefont
  {He}}, \bibinfo {author} {\bibfnamefont {Y.}~\bibnamefont {Wang}}, \bibinfo
  {author} {\bibfnamefont {L.}~\bibnamefont {Lu}}, \bibinfo {author}
  {\bibfnamefont {X.-C.}\ \bibnamefont {Ma}}, \ and\ \bibinfo {author}
  {\bibfnamefont {Q.-K.}\ \bibnamefont {Xue}},\ }\href {\doibase
  10.1126/science.1234414} {\bibfield  {journal} {\bibinfo  {journal}
  {Science}\ }\textbf {\bibinfo {volume} {340}},\ \bibinfo {pages} {167}
  (\bibinfo {year} {2013})}\BibitemShut {NoStop}%
\bibitem [{\citenamefont {Hirahara}\ \emph {et~al.}(2017)\citenamefont
  {Hirahara}, \citenamefont {Eremeev}, \citenamefont {Shirasawa}, \citenamefont
  {Okuyama}, \citenamefont {Kubo}, \citenamefont {Nakanishi}, \citenamefont
  {Akiyama}, \citenamefont {Takayama}, \citenamefont {Hajiri}, \citenamefont
  {Ideta}, \citenamefont {Matsunami}, \citenamefont {Sumida}, \citenamefont
  {Miyamoto}, \citenamefont {Takagi}, \citenamefont {Tanaka}, \citenamefont
  {Okuda}, \citenamefont {Yokoyama}, \citenamefont {Kimura}, \citenamefont
  {Hasegawa},\ and\ \citenamefont {Chulkov}}]{Hirahara17}%
  \BibitemOpen
  \bibfield  {author} {\bibinfo {author} {\bibfnamefont {T.}~\bibnamefont
  {Hirahara}}, \bibinfo {author} {\bibfnamefont {S.~V.}\ \bibnamefont
  {Eremeev}}, \bibinfo {author} {\bibfnamefont {T.}~\bibnamefont {Shirasawa}},
  \bibinfo {author} {\bibfnamefont {Y.}~\bibnamefont {Okuyama}}, \bibinfo
  {author} {\bibfnamefont {T.}~\bibnamefont {Kubo}}, \bibinfo {author}
  {\bibfnamefont {R.}~\bibnamefont {Nakanishi}}, \bibinfo {author}
  {\bibfnamefont {R.}~\bibnamefont {Akiyama}}, \bibinfo {author} {\bibfnamefont
  {A.}~\bibnamefont {Takayama}}, \bibinfo {author} {\bibfnamefont
  {T.}~\bibnamefont {Hajiri}}, \bibinfo {author} {\bibfnamefont {S.-i.}\
  \bibnamefont {Ideta}}, \bibinfo {author} {\bibfnamefont {M.}~\bibnamefont
  {Matsunami}}, \bibinfo {author} {\bibfnamefont {K.}~\bibnamefont {Sumida}},
  \bibinfo {author} {\bibfnamefont {K.}~\bibnamefont {Miyamoto}}, \bibinfo
  {author} {\bibfnamefont {Y.}~\bibnamefont {Takagi}}, \bibinfo {author}
  {\bibfnamefont {K.}~\bibnamefont {Tanaka}}, \bibinfo {author} {\bibfnamefont
  {T.}~\bibnamefont {Okuda}}, \bibinfo {author} {\bibfnamefont
  {T.}~\bibnamefont {Yokoyama}}, \bibinfo {author} {\bibfnamefont {S.-i.}\
  \bibnamefont {Kimura}}, \bibinfo {author} {\bibfnamefont {S.}~\bibnamefont
  {Hasegawa}}, \ and\ \bibinfo {author} {\bibfnamefont {E.~V.}\ \bibnamefont
  {Chulkov}},\ }\href {http://dx.doi.org/10.1021/acs.nanolett.7b00560}
  {\bibfield  {journal} {\bibinfo  {journal} {Nano Letters}\ }\textbf {\bibinfo
  {volume} {0}} (\bibinfo {year} {2017})}\BibitemShut {NoStop}%
\bibitem [{\citenamefont {Mogi}\ \emph {et~al.}(2017)\citenamefont {Mogi},
  \citenamefont {Kawamura}, \citenamefont {Yoshimi}, \citenamefont {Tsukazaki},
  \citenamefont {Kozuka}, \citenamefont {Shirakawa}, \citenamefont {Takahashi},
  \citenamefont {Kawasaki},\ and\ \citenamefont {Tokura}}]{Mogi17}%
  \BibitemOpen
  \bibfield  {author} {\bibinfo {author} {\bibfnamefont {M.}~\bibnamefont
  {Mogi}}, \bibinfo {author} {\bibfnamefont {M.}~\bibnamefont {Kawamura}},
  \bibinfo {author} {\bibfnamefont {R.}~\bibnamefont {Yoshimi}}, \bibinfo
  {author} {\bibfnamefont {A.}~\bibnamefont {Tsukazaki}}, \bibinfo {author}
  {\bibfnamefont {Y.}~\bibnamefont {Kozuka}}, \bibinfo {author} {\bibfnamefont
  {N.}~\bibnamefont {Shirakawa}}, \bibinfo {author} {\bibfnamefont {K.~S.}\
  \bibnamefont {Takahashi}}, \bibinfo {author} {\bibfnamefont {M.}~\bibnamefont
  {Kawasaki}}, \ and\ \bibinfo {author} {\bibfnamefont {Y.}~\bibnamefont
  {Tokura}},\ }\href {http://dx.doi.org/10.1038/nmat4855} {\bibfield  {journal}
  {\bibinfo  {journal} {Nat Mater}\ ,\ \bibinfo {pages} {4855}} (\bibinfo
  {year} {2017})}\BibitemShut {NoStop}%
\bibitem [{\citenamefont {He}\ \emph {et~al.}(2017)\citenamefont {He},
  \citenamefont {Kou}, \citenamefont {Grutter}, \citenamefont {Yin},
  \citenamefont {Pan}, \citenamefont {Che}, \citenamefont {Liu}, \citenamefont
  {Nie}, \citenamefont {Zhang}, \citenamefont {Disseler}, \citenamefont
  {Kirby}, \citenamefont {{Ratcliff II}}, \citenamefont {Shao}, \citenamefont
  {Murata}, \citenamefont {Zhu}, \citenamefont {Yu}, \citenamefont {Fan},
  \citenamefont {Montazeri}, \citenamefont {Han}, \citenamefont {Borchers},\
  and\ \citenamefont {Wang}}]{Q-L-He17}%
  \BibitemOpen
  \bibfield  {author} {\bibinfo {author} {\bibfnamefont {Q.~L.}\ \bibnamefont
  {He}}, \bibinfo {author} {\bibfnamefont {X.}~\bibnamefont {Kou}}, \bibinfo
  {author} {\bibfnamefont {A.~J.}\ \bibnamefont {Grutter}}, \bibinfo {author}
  {\bibfnamefont {G.}~\bibnamefont {Yin}}, \bibinfo {author} {\bibfnamefont
  {L.}~\bibnamefont {Pan}}, \bibinfo {author} {\bibfnamefont {X.}~\bibnamefont
  {Che}}, \bibinfo {author} {\bibfnamefont {Y.}~\bibnamefont {Liu}}, \bibinfo
  {author} {\bibfnamefont {T.}~\bibnamefont {Nie}}, \bibinfo {author}
  {\bibfnamefont {B.}~\bibnamefont {Zhang}}, \bibinfo {author} {\bibfnamefont
  {S.~M.}\ \bibnamefont {Disseler}}, \bibinfo {author} {\bibfnamefont {B.~J.}\
  \bibnamefont {Kirby}}, \bibinfo {author} {\bibfnamefont {W.}~\bibnamefont
  {{Ratcliff II}}}, \bibinfo {author} {\bibfnamefont {Q.}~\bibnamefont {Shao}},
  \bibinfo {author} {\bibfnamefont {K.}~\bibnamefont {Murata}}, \bibinfo
  {author} {\bibfnamefont {X.}~\bibnamefont {Zhu}}, \bibinfo {author}
  {\bibfnamefont {G.}~\bibnamefont {Yu}}, \bibinfo {author} {\bibfnamefont
  {Y.}~\bibnamefont {Fan}}, \bibinfo {author} {\bibfnamefont {M.}~\bibnamefont
  {Montazeri}}, \bibinfo {author} {\bibfnamefont {X.}~\bibnamefont {Han}},
  \bibinfo {author} {\bibfnamefont {J.~A.}\ \bibnamefont {Borchers}}, \ and\
  \bibinfo {author} {\bibfnamefont {K.~L.}\ \bibnamefont {Wang}},\ }\href
  {http://dx.doi.org/10.1038/nmat4783 http://10.0.4.14/nmat4783
  http://www.nature.com/nmat/journal/v16/n1/abs/nmat4783.html{\#}supplementary-information}
  {\bibfield  {journal} {\bibinfo  {journal} {Nat Mater}\ }\textbf {\bibinfo
  {volume} {16}},\ \bibinfo {pages} {94} (\bibinfo {year} {2017})}\BibitemShut
  {NoStop}%
\bibitem [{\citenamefont {Huang}\ \emph {et~al.}(2015)\citenamefont {Huang},
  \citenamefont {Xu}, \citenamefont {Belopolski}, \citenamefont {Lee},
  \citenamefont {Chang}, \citenamefont {Wang}, \citenamefont {Alidoust},
  \citenamefont {Bian}, \citenamefont {Neupane}, \citenamefont {Zhang},
  \citenamefont {Jia}, \citenamefont {Bansil}, \citenamefont {Lin},\ and\
  \citenamefont {Hasan}}]{Huang15}%
  \BibitemOpen
  \bibfield  {author} {\bibinfo {author} {\bibfnamefont {S.-M.}\ \bibnamefont
  {Huang}}, \bibinfo {author} {\bibfnamefont {S.-Y.}\ \bibnamefont {Xu}},
  \bibinfo {author} {\bibfnamefont {I.}~\bibnamefont {Belopolski}}, \bibinfo
  {author} {\bibfnamefont {C.-C.}\ \bibnamefont {Lee}}, \bibinfo {author}
  {\bibfnamefont {G.}~\bibnamefont {Chang}}, \bibinfo {author} {\bibfnamefont
  {B.}~\bibnamefont {Wang}}, \bibinfo {author} {\bibfnamefont {N.}~\bibnamefont
  {Alidoust}}, \bibinfo {author} {\bibfnamefont {G.}~\bibnamefont {Bian}},
  \bibinfo {author} {\bibfnamefont {M.}~\bibnamefont {Neupane}}, \bibinfo
  {author} {\bibfnamefont {C.}~\bibnamefont {Zhang}}, \bibinfo {author}
  {\bibfnamefont {S.}~\bibnamefont {Jia}}, \bibinfo {author} {\bibfnamefont
  {A.}~\bibnamefont {Bansil}}, \bibinfo {author} {\bibfnamefont
  {H.}~\bibnamefont {Lin}}, \ and\ \bibinfo {author} {\bibfnamefont {M.~Z.}\
  \bibnamefont {Hasan}},\ }\href {\doibase 10.1038/ncomms8373} {\bibfield
  {journal} {\bibinfo  {journal} {Nat. Commun.}\ }\textbf {\bibinfo {volume}
  {6}},\ \bibinfo {pages} {7373} (\bibinfo {year} {2015})}\BibitemShut
  {NoStop}%
\bibitem [{\citenamefont {Xu}\ \emph {et~al.}(2015)\citenamefont {Xu},
  \citenamefont {Belopolski}, \citenamefont {Alidoust}, \citenamefont
  {Neupane}, \citenamefont {Bian}, \citenamefont {Zhang}, \citenamefont
  {Sankar}, \citenamefont {Chang}, \citenamefont {Yuan}, \citenamefont {Lee},
  \citenamefont {Huang}, \citenamefont {Zheng}, \citenamefont {Ma},
  \citenamefont {Sanchez}, \citenamefont {Wang}, \citenamefont {Bansil},
  \citenamefont {Chou}, \citenamefont {Shibayev}, \citenamefont {Lin},
  \citenamefont {Jia},\ and\ \citenamefont {Hasan}}]{S.Xu15}%
  \BibitemOpen
  \bibfield  {author} {\bibinfo {author} {\bibfnamefont {S.-Y.}\ \bibnamefont
  {Xu}}, \bibinfo {author} {\bibfnamefont {I.}~\bibnamefont {Belopolski}},
  \bibinfo {author} {\bibfnamefont {N.}~\bibnamefont {Alidoust}}, \bibinfo
  {author} {\bibfnamefont {M.}~\bibnamefont {Neupane}}, \bibinfo {author}
  {\bibfnamefont {G.}~\bibnamefont {Bian}}, \bibinfo {author} {\bibfnamefont
  {C.}~\bibnamefont {Zhang}}, \bibinfo {author} {\bibfnamefont
  {R.}~\bibnamefont {Sankar}}, \bibinfo {author} {\bibfnamefont
  {G.}~\bibnamefont {Chang}}, \bibinfo {author} {\bibfnamefont
  {Z.}~\bibnamefont {Yuan}}, \bibinfo {author} {\bibfnamefont {C.-C.}\
  \bibnamefont {Lee}}, \bibinfo {author} {\bibfnamefont {S.-M.}\ \bibnamefont
  {Huang}}, \bibinfo {author} {\bibfnamefont {H.}~\bibnamefont {Zheng}},
  \bibinfo {author} {\bibfnamefont {J.}~\bibnamefont {Ma}}, \bibinfo {author}
  {\bibfnamefont {D.~S.}\ \bibnamefont {Sanchez}}, \bibinfo {author}
  {\bibfnamefont {B.}~\bibnamefont {Wang}}, \bibinfo {author} {\bibfnamefont
  {A.}~\bibnamefont {Bansil}}, \bibinfo {author} {\bibfnamefont
  {F.}~\bibnamefont {Chou}}, \bibinfo {author} {\bibfnamefont {P.~P.}\
  \bibnamefont {Shibayev}}, \bibinfo {author} {\bibfnamefont {H.}~\bibnamefont
  {Lin}}, \bibinfo {author} {\bibfnamefont {S.}~\bibnamefont {Jia}}, \ and\
  \bibinfo {author} {\bibfnamefont {M.~Z.}\ \bibnamefont {Hasan}},\ }\href
  {\doibase 10.1126/science.aaa9297} {\bibfield  {journal} {\bibinfo  {journal}
  {Science}\ }\textbf {\bibinfo {volume} {349}},\ \bibinfo {pages} {613}
  (\bibinfo {year} {2015})}\BibitemShut {NoStop}%
\bibitem [{\citenamefont {Lv}\ \emph {et~al.}(2015)\citenamefont {Lv},
  \citenamefont {Weng}, \citenamefont {Fu}, \citenamefont {Wang}, \citenamefont
  {Miao}, \citenamefont {Ma}, \citenamefont {Richard}, \citenamefont {Huang},
  \citenamefont {Zhao}, \citenamefont {Chen}, \citenamefont {Fang},
  \citenamefont {Dai}, \citenamefont {Qian},\ and\ \citenamefont
  {Ding}}]{Lv15}%
  \BibitemOpen
  \bibfield  {author} {\bibinfo {author} {\bibfnamefont {B.~Q.}\ \bibnamefont
  {Lv}}, \bibinfo {author} {\bibfnamefont {H.~M.}\ \bibnamefont {Weng}},
  \bibinfo {author} {\bibfnamefont {B.~B.}\ \bibnamefont {Fu}}, \bibinfo
  {author} {\bibfnamefont {X.~P.}\ \bibnamefont {Wang}}, \bibinfo {author}
  {\bibfnamefont {H.}~\bibnamefont {Miao}}, \bibinfo {author} {\bibfnamefont
  {J.}~\bibnamefont {Ma}}, \bibinfo {author} {\bibfnamefont {P.}~\bibnamefont
  {Richard}}, \bibinfo {author} {\bibfnamefont {X.~C.}\ \bibnamefont {Huang}},
  \bibinfo {author} {\bibfnamefont {L.~X.}\ \bibnamefont {Zhao}}, \bibinfo
  {author} {\bibfnamefont {G.~F.}\ \bibnamefont {Chen}}, \bibinfo {author}
  {\bibfnamefont {Z.}~\bibnamefont {Fang}}, \bibinfo {author} {\bibfnamefont
  {X.}~\bibnamefont {Dai}}, \bibinfo {author} {\bibfnamefont {T.}~\bibnamefont
  {Qian}}, \ and\ \bibinfo {author} {\bibfnamefont {H.}~\bibnamefont {Ding}},\
  }\href {\doibase 10.1103/PhysRevX.5.031013} {\bibfield  {journal} {\bibinfo
  {journal} {Phys. Rev. X}\ }\textbf {\bibinfo {volume} {5}},\ \bibinfo {pages}
  {031013} (\bibinfo {year} {2015})}\BibitemShut {NoStop}%
\bibitem [{\citenamefont {Weng}\ \emph {et~al.}(2015)\citenamefont {Weng},
  \citenamefont {Fang}, \citenamefont {Fang}, \citenamefont {Bernevig},\ and\
  \citenamefont {Dai}}]{Weng15}%
  \BibitemOpen
  \bibfield  {author} {\bibinfo {author} {\bibfnamefont {H.}~\bibnamefont
  {Weng}}, \bibinfo {author} {\bibfnamefont {C.}~\bibnamefont {Fang}}, \bibinfo
  {author} {\bibfnamefont {Z.}~\bibnamefont {Fang}}, \bibinfo {author}
  {\bibfnamefont {B.~A.}\ \bibnamefont {Bernevig}}, \ and\ \bibinfo {author}
  {\bibfnamefont {X.}~\bibnamefont {Dai}},\ }\href {\doibase
  10.1103/PhysRevX.5.011029} {\bibfield  {journal} {\bibinfo  {journal} {Phys.
  Rev. X}\ }\textbf {\bibinfo {volume} {5}},\ \bibinfo {pages} {011029}
  (\bibinfo {year} {2015})}\BibitemShut {NoStop}%
\bibitem [{\citenamefont {Burkov}\ and\ \citenamefont
  {Balents}(2011)}]{Burkov11}%
  \BibitemOpen
  \bibfield  {author} {\bibinfo {author} {\bibfnamefont {A.~A.}\ \bibnamefont
  {Burkov}}\ and\ \bibinfo {author} {\bibfnamefont {L.}~\bibnamefont
  {Balents}},\ }\href {\doibase 10.1103/PhysRevLett.107.127205} {\bibfield
  {journal} {\bibinfo  {journal} {Phys. Rev. Lett.}\ }\textbf {\bibinfo
  {volume} {107}},\ \bibinfo {pages} {127205} (\bibinfo {year}
  {2011})}\BibitemShut {NoStop}%
\bibitem [{\citenamefont {Tominaga}\ \emph {et~al.}(2014)\citenamefont
  {Tominaga}, \citenamefont {Kolobov}, \citenamefont {Fons}, \citenamefont
  {Nakano},\ and\ \citenamefont {Murakami}}]{Tominaga14}%
  \BibitemOpen
  \bibfield  {author} {\bibinfo {author} {\bibfnamefont {J.}~\bibnamefont
  {Tominaga}}, \bibinfo {author} {\bibfnamefont {a.~V.}\ \bibnamefont
  {Kolobov}}, \bibinfo {author} {\bibfnamefont {P.}~\bibnamefont {Fons}},
  \bibinfo {author} {\bibfnamefont {T.}~\bibnamefont {Nakano}}, \ and\ \bibinfo
  {author} {\bibfnamefont {S.}~\bibnamefont {Murakami}},\ }\href {\doibase
  10.1002/admi.201300027} {\bibfield  {journal} {\bibinfo  {journal} {Adv.
  Mater. Interfaces}\ }\textbf {\bibinfo {volume} {1}},\ \bibinfo {pages}
  {1300027} (\bibinfo {year} {2014})}\BibitemShut {NoStop}%
\bibitem [{\citenamefont {Tominaga}\ \emph {et~al.}(2015)\citenamefont
  {Tominaga}, \citenamefont {Kolobov}, \citenamefont {Fons}, \citenamefont
  {Wang}, \citenamefont {Saito}, \citenamefont {Nakano}, \citenamefont {Hase},
  \citenamefont {Murakami}, \citenamefont {Herfort},\ and\ \citenamefont
  {Takagaki}}]{Tominaga15}%
  \BibitemOpen
  \bibfield  {author} {\bibinfo {author} {\bibfnamefont {J.}~\bibnamefont
  {Tominaga}}, \bibinfo {author} {\bibfnamefont {A.~V.}\ \bibnamefont
  {Kolobov}}, \bibinfo {author} {\bibfnamefont {P.~J.}\ \bibnamefont {Fons}},
  \bibinfo {author} {\bibfnamefont {X.}~\bibnamefont {Wang}}, \bibinfo {author}
  {\bibfnamefont {Y.}~\bibnamefont {Saito}}, \bibinfo {author} {\bibfnamefont
  {T.}~\bibnamefont {Nakano}}, \bibinfo {author} {\bibfnamefont
  {M.}~\bibnamefont {Hase}}, \bibinfo {author} {\bibfnamefont {S.}~\bibnamefont
  {Murakami}}, \bibinfo {author} {\bibfnamefont {J.}~\bibnamefont {Herfort}}, \
  and\ \bibinfo {author} {\bibfnamefont {Y.}~\bibnamefont {Takagaki}},\ }\href
  {http://stacks.iop.org/1468-6996/16/i=1/a=014402} {\bibfield  {journal}
  {\bibinfo  {journal} {Sci. Technol. Adv. Mater.}\ }\textbf {\bibinfo {volume}
  {16}},\ \bibinfo {pages} {14402} (\bibinfo {year} {2015})}\BibitemShut
  {NoStop}%
\bibitem [{\citenamefont {Essin}\ \emph {et~al.}(2009)\citenamefont {Essin},
  \citenamefont {Moore},\ and\ \citenamefont {Vanderbilt}}]{Essin09}%
  \BibitemOpen
  \bibfield  {author} {\bibinfo {author} {\bibfnamefont {A.~M.}\ \bibnamefont
  {Essin}}, \bibinfo {author} {\bibfnamefont {J.~E.}\ \bibnamefont {Moore}}, \
  and\ \bibinfo {author} {\bibfnamefont {D.}~\bibnamefont {Vanderbilt}},\
  }\href {\doibase 10.1103/PhysRevLett.102.146805} {\bibfield  {journal}
  {\bibinfo  {journal} {Phys. Rev. Lett.}\ }\textbf {\bibinfo {volume} {102}},\
  \bibinfo {pages} {146805} (\bibinfo {year} {2009})}\BibitemShut {NoStop}%
\bibitem [{\citenamefont {Qi}\ \emph {et~al.}(2008)\citenamefont {Qi},
  \citenamefont {Hughes},\ and\ \citenamefont {Zhang}}]{Qi08a}%
  \BibitemOpen
  \bibfield  {author} {\bibinfo {author} {\bibfnamefont {X.-L.}\ \bibnamefont
  {Qi}}, \bibinfo {author} {\bibfnamefont {T.~L.}\ \bibnamefont {Hughes}}, \
  and\ \bibinfo {author} {\bibfnamefont {S.-C.}\ \bibnamefont {Zhang}},\ }\href
  {\doibase 10.1103/PhysRevB.78.195424} {\bibfield  {journal} {\bibinfo
  {journal} {Phys. Rev. B}\ }\textbf {\bibinfo {volume} {78}},\ \bibinfo
  {pages} {195424} (\bibinfo {year} {2008})}\BibitemShut {NoStop}%
\bibitem [{\citenamefont {Qi}\ \emph {et~al.}(2009)\citenamefont {Qi},
  \citenamefont {Li}, \citenamefont {Zang},\ and\ \citenamefont
  {Zhang}}]{Qi09a}%
  \BibitemOpen
  \bibfield  {author} {\bibinfo {author} {\bibfnamefont {X.-L.}\ \bibnamefont
  {Qi}}, \bibinfo {author} {\bibfnamefont {R.}~\bibnamefont {Li}}, \bibinfo
  {author} {\bibfnamefont {J.}~\bibnamefont {Zang}}, \ and\ \bibinfo {author}
  {\bibfnamefont {S.-C.}\ \bibnamefont {Zhang}},\ }\href {\doibase
  10.1126/science.1167747} {\bibfield  {journal} {\bibinfo  {journal}
  {Science}\ }\textbf {\bibinfo {volume} {323}},\ \bibinfo {pages} {1184}
  (\bibinfo {year} {2009})}\BibitemShut {NoStop}%
\bibitem [{\citenamefont {Karch}(2009)}]{Karch09}%
  \BibitemOpen
  \bibfield  {author} {\bibinfo {author} {\bibfnamefont {A.}~\bibnamefont
  {Karch}},\ }\href {\doibase 10.1103/PhysRevLett.103.171601} {\bibfield
  {journal} {\bibinfo  {journal} {Phys. Rev. Lett.}\ }\textbf {\bibinfo
  {volume} {103}},\ \bibinfo {pages} {171601} (\bibinfo {year}
  {2009})}\BibitemShut {NoStop}%
\bibitem [{\citenamefont {Rosenberg}\ and\ \citenamefont
  {Franz}(2010)}]{Rosenberg10}%
  \BibitemOpen
  \bibfield  {author} {\bibinfo {author} {\bibfnamefont {G.}~\bibnamefont
  {Rosenberg}}\ and\ \bibinfo {author} {\bibfnamefont {M.}~\bibnamefont
  {Franz}},\ }\href {\doibase 10.1103/PhysRevB.82.035105} {\bibfield  {journal}
  {\bibinfo  {journal} {Phys. Rev. B}\ }\textbf {\bibinfo {volume} {82}},\
  \bibinfo {pages} {035105} (\bibinfo {year} {2010})}\BibitemShut {NoStop}%
\bibitem [{\citenamefont {Vazifeh}\ and\ \citenamefont
  {Franz}(2010)}]{Vazifeh10}%
  \BibitemOpen
  \bibfield  {author} {\bibinfo {author} {\bibfnamefont {M.~M.}\ \bibnamefont
  {Vazifeh}}\ and\ \bibinfo {author} {\bibfnamefont {M.}~\bibnamefont
  {Franz}},\ }\href {\doibase 10.1103/PhysRevB.82.233103} {\bibfield  {journal}
  {\bibinfo  {journal} {Phys. Rev. B}\ }\textbf {\bibinfo {volume} {82}},\
  \bibinfo {pages} {233103} (\bibinfo {year} {2010})}\BibitemShut {NoStop}%
\bibitem [{\citenamefont {Lan}\ \emph {et~al.}(2011)\citenamefont {Lan},
  \citenamefont {Wan},\ and\ \citenamefont {Zhang}}]{Lan11}%
  \BibitemOpen
  \bibfield  {author} {\bibinfo {author} {\bibfnamefont {Y.}~\bibnamefont
  {Lan}}, \bibinfo {author} {\bibfnamefont {S.}~\bibnamefont {Wan}}, \ and\
  \bibinfo {author} {\bibfnamefont {S.-C.}\ \bibnamefont {Zhang}},\ }\href
  {\doibase 10.1103/PhysRevB.83.205109} {\bibfield  {journal} {\bibinfo
  {journal} {Phys. Rev. B}\ }\textbf {\bibinfo {volume} {83}},\ \bibinfo
  {pages} {205109} (\bibinfo {year} {2011})}\BibitemShut {NoStop}%
\bibitem [{\citenamefont {Yu}\ \emph {et~al.}(2010)\citenamefont {Yu},
  \citenamefont {Zhang}, \citenamefont {Zhang}, \citenamefont {Zhang},
  \citenamefont {Dai},\ and\ \citenamefont {Fang}}]{R-Yu10}%
  \BibitemOpen
  \bibfield  {author} {\bibinfo {author} {\bibfnamefont {R.}~\bibnamefont
  {Yu}}, \bibinfo {author} {\bibfnamefont {W.}~\bibnamefont {Zhang}}, \bibinfo
  {author} {\bibfnamefont {H.-J.}\ \bibnamefont {Zhang}}, \bibinfo {author}
  {\bibfnamefont {S.-C.}\ \bibnamefont {Zhang}}, \bibinfo {author}
  {\bibfnamefont {X.}~\bibnamefont {Dai}}, \ and\ \bibinfo {author}
  {\bibfnamefont {Z.}~\bibnamefont {Fang}},\ }\href {\doibase
  10.1126/science.1187485} {\bibfield  {journal} {\bibinfo  {journal}
  {Science}\ }\textbf {\bibinfo {volume} {329}},\ \bibinfo {pages} {61}
  (\bibinfo {year} {2010})}\BibitemShut {NoStop}%
\bibitem [{\citenamefont {Nomura}\ and\ \citenamefont
  {Nagaosa}(2011)}]{Nomura11}%
  \BibitemOpen
  \bibfield  {author} {\bibinfo {author} {\bibfnamefont {K.}~\bibnamefont
  {Nomura}}\ and\ \bibinfo {author} {\bibfnamefont {N.}~\bibnamefont
  {Nagaosa}},\ }\href {\doibase 10.1103/PhysRevLett.106.166802} {\bibfield
  {journal} {\bibinfo  {journal} {Phys. Rev. Lett.}\ }\textbf {\bibinfo
  {volume} {106}},\ \bibinfo {pages} {166802} (\bibinfo {year}
  {2011})}\BibitemShut {NoStop}%
\bibitem [{\citenamefont {Wang}\ \emph {et~al.}(2015)\citenamefont {Wang},
  \citenamefont {Lian}, \citenamefont {Qi},\ and\ \citenamefont
  {Zhang}}]{J-Wang15}%
  \BibitemOpen
  \bibfield  {author} {\bibinfo {author} {\bibfnamefont {J.}~\bibnamefont
  {Wang}}, \bibinfo {author} {\bibfnamefont {B.}~\bibnamefont {Lian}}, \bibinfo
  {author} {\bibfnamefont {X.-L.}\ \bibnamefont {Qi}}, \ and\ \bibinfo {author}
  {\bibfnamefont {S.-C.}\ \bibnamefont {Zhang}},\ }\href {\doibase
  10.1103/PhysRevB.92.081107} {\bibfield  {journal} {\bibinfo  {journal} {Phys.
  Rev. B}\ }\textbf {\bibinfo {volume} {92}},\ \bibinfo {pages} {081107}
  (\bibinfo {year} {2015})}\BibitemShut {NoStop}%
\bibitem [{\citenamefont {Zyuzin}\ and\ \citenamefont
  {Burkov}(2012)}]{Zyuzin12a}%
  \BibitemOpen
  \bibfield  {author} {\bibinfo {author} {\bibfnamefont {A.~A.}\ \bibnamefont
  {Zyuzin}}\ and\ \bibinfo {author} {\bibfnamefont {A.~A.}\ \bibnamefont
  {Burkov}},\ }\href {\doibase 10.1103/PhysRevB.86.115133} {\bibfield
  {journal} {\bibinfo  {journal} {Phys. Rev. B}\ }\textbf {\bibinfo {volume}
  {86}},\ \bibinfo {pages} {1} (\bibinfo {year} {2012})}\BibitemShut {NoStop}%
\bibitem [{\citenamefont {Okada}\ \emph {et~al.}(2016)\citenamefont {Okada},
  \citenamefont {Takahashi}, \citenamefont {Mogi}, \citenamefont {Yoshimi},
  \citenamefont {Tsukazaki}, \citenamefont {Takahashi}, \citenamefont {Ogawa},
  \citenamefont {Kawasaki},\ and\ \citenamefont {Tokura}}]{Okada16}%
  \BibitemOpen
  \bibfield  {author} {\bibinfo {author} {\bibfnamefont {K.~N.}\ \bibnamefont
  {Okada}}, \bibinfo {author} {\bibfnamefont {Y.}~\bibnamefont {Takahashi}},
  \bibinfo {author} {\bibfnamefont {M.}~\bibnamefont {Mogi}}, \bibinfo {author}
  {\bibfnamefont {R.}~\bibnamefont {Yoshimi}}, \bibinfo {author} {\bibfnamefont
  {A.}~\bibnamefont {Tsukazaki}}, \bibinfo {author} {\bibfnamefont {K.~S.}\
  \bibnamefont {Takahashi}}, \bibinfo {author} {\bibfnamefont {N.}~\bibnamefont
  {Ogawa}}, \bibinfo {author} {\bibfnamefont {M.}~\bibnamefont {Kawasaki}}, \
  and\ \bibinfo {author} {\bibfnamefont {Y.}~\bibnamefont {Tokura}},\ }\href
  {http://dx.doi.org/10.1038/ncomms12245 http://10.0.4.14/ncomms12245
  http://www.nature.com/articles/ncomms12245{\#}supplementary-information}
  {\bibfield  {journal} {\bibinfo  {journal} {Nat. Commun.}\ }\textbf {\bibinfo
  {volume} {7}},\ \bibinfo {pages} {12245} (\bibinfo {year}
  {2016})}\BibitemShut {NoStop}%
\bibitem [{\citenamefont {Vilenkin}(1980)}]{Vilenkin80}%
  \BibitemOpen
  \bibfield  {author} {\bibinfo {author} {\bibfnamefont {A.}~\bibnamefont
  {Vilenkin}},\ }\href {\doibase 10.1103/PhysRevD.22.3080} {\bibfield
  {journal} {\bibinfo  {journal} {Phys. Rev. D}\ }\textbf {\bibinfo {volume}
  {22}},\ \bibinfo {pages} {3080} (\bibinfo {year} {1980})}\BibitemShut
  {NoStop}%
\bibitem [{\citenamefont {Kharzeev}\ \emph {et~al.}(2008)\citenamefont
  {Kharzeev}, \citenamefont {McLerran},\ and\ \citenamefont
  {Warringa}}]{Kharzeev08}%
  \BibitemOpen
  \bibfield  {author} {\bibinfo {author} {\bibfnamefont {D.~E.}\ \bibnamefont
  {Kharzeev}}, \bibinfo {author} {\bibfnamefont {L.~D.}\ \bibnamefont
  {McLerran}}, \ and\ \bibinfo {author} {\bibfnamefont {H.~J.}\ \bibnamefont
  {Warringa}},\ }\href {\doibase
  http://dx.doi.org/10.1016/j.nuclphysa.2008.02.298} {\bibfield  {journal}
  {\bibinfo  {journal} {Nuclear Physics A}\ }\textbf {\bibinfo {volume}
  {803}},\ \bibinfo {pages} {227 } (\bibinfo {year} {2008})}\BibitemShut
  {NoStop}%
\bibitem [{\citenamefont {Fukushima}\ \emph {et~al.}(2008)\citenamefont
  {Fukushima}, \citenamefont {Kharzeev},\ and\ \citenamefont
  {Warringa}}]{Fukushima08}%
  \BibitemOpen
  \bibfield  {author} {\bibinfo {author} {\bibfnamefont {K.}~\bibnamefont
  {Fukushima}}, \bibinfo {author} {\bibfnamefont {D.~E.}\ \bibnamefont
  {Kharzeev}}, \ and\ \bibinfo {author} {\bibfnamefont {H.~J.}\ \bibnamefont
  {Warringa}},\ }\href {\doibase 10.1103/PhysRevD.78.074033} {\bibfield
  {journal} {\bibinfo  {journal} {Phys. Rev. D}\ }\textbf {\bibinfo {volume}
  {78}},\ \bibinfo {pages} {074033} (\bibinfo {year} {2008})}\BibitemShut
  {NoStop}%
\bibitem [{\citenamefont {Vazifeh}\ and\ \citenamefont
  {Franz}(2013)}]{Vazifeh13}%
  \BibitemOpen
  \bibfield  {author} {\bibinfo {author} {\bibfnamefont {M.~M.}\ \bibnamefont
  {Vazifeh}}\ and\ \bibinfo {author} {\bibfnamefont {M.}~\bibnamefont
  {Franz}},\ }\href {\doibase 10.1103/PhysRevLett.111.027201} {\bibfield
  {journal} {\bibinfo  {journal} {Phys. Rev. Lett.}\ }\textbf {\bibinfo
  {volume} {111}},\ \bibinfo {pages} {027201} (\bibinfo {year}
  {2013})}\BibitemShut {NoStop}%
\bibitem [{\citenamefont {Sumiyoshi}\ and\ \citenamefont
  {Fujimoto}(2016)}]{Sumiyoshi16}%
  \BibitemOpen
  \bibfield  {author} {\bibinfo {author} {\bibfnamefont {H.}~\bibnamefont
  {Sumiyoshi}}\ and\ \bibinfo {author} {\bibfnamefont {S.}~\bibnamefont
  {Fujimoto}},\ }\href {\doibase 10.1103/PhysRevLett.116.166601} {\bibfield
  {journal} {\bibinfo  {journal} {Phys. Rev. Lett.}\ }\textbf {\bibinfo
  {volume} {116}},\ \bibinfo {pages} {166601} (\bibinfo {year}
  {2016})}\BibitemShut {NoStop}%
\bibitem [{\citenamefont {Sekine}\ and\ \citenamefont
  {Nomura}(2016)}]{Sekine16}%
  \BibitemOpen
  \bibfield  {author} {\bibinfo {author} {\bibfnamefont {A.}~\bibnamefont
  {Sekine}}\ and\ \bibinfo {author} {\bibfnamefont {K.}~\bibnamefont
  {Nomura}},\ }\href {\doibase 10.1103/PhysRevLett.116.096401} {\bibfield
  {journal} {\bibinfo  {journal} {Phys. Rev. Lett.}\ }\textbf {\bibinfo
  {volume} {116}},\ \bibinfo {pages} {96401} (\bibinfo {year}
  {2016})}\BibitemShut {NoStop}%
\bibitem [{\citenamefont {Taguchi}\ \emph {et~al.}(2016)\citenamefont
  {Taguchi}, \citenamefont {Imaeda}, \citenamefont {Sato},\ and\ \citenamefont
  {Tanaka}}]{Taguchi16a}%
  \BibitemOpen
  \bibfield  {author} {\bibinfo {author} {\bibfnamefont {K.}~\bibnamefont
  {Taguchi}}, \bibinfo {author} {\bibfnamefont {T.}~\bibnamefont {Imaeda}},
  \bibinfo {author} {\bibfnamefont {M.}~\bibnamefont {Sato}}, \ and\ \bibinfo
  {author} {\bibfnamefont {Y.}~\bibnamefont {Tanaka}},\ }\href {\doibase
  10.1103/PhysRevB.93.201202} {\bibfield  {journal} {\bibinfo  {journal} {Phys.
  Rev. B}\ }\textbf {\bibinfo {volume} {93}},\ \bibinfo {pages} {201202}
  (\bibinfo {year} {2016})}\BibitemShut {NoStop}%
\bibitem [{\citenamefont {Li}\ \emph {et~al.}(2016)\citenamefont {Li},
  \citenamefont {Kharzeev}, \citenamefont {Zhang}, \citenamefont {Huang},
  \citenamefont {Pletikosic}, \citenamefont {Fedorov}, \citenamefont {Zhong},
  \citenamefont {Schneeloch}, \citenamefont {Gu}, \citenamefont {Valla},
  \citenamefont {Pletikosi{\'{c}}}, \citenamefont {Fedorov}, \citenamefont
  {Zhong}, \citenamefont {Schneeloch}, \citenamefont {Gu},\ and\ \citenamefont
  {Valla}}]{Li2016}%
  \BibitemOpen
  \bibfield  {author} {\bibinfo {author} {\bibfnamefont {Q.}~\bibnamefont
  {Li}}, \bibinfo {author} {\bibfnamefont {D.~E.}\ \bibnamefont {Kharzeev}},
  \bibinfo {author} {\bibfnamefont {C.}~\bibnamefont {Zhang}}, \bibinfo
  {author} {\bibfnamefont {Y.}~\bibnamefont {Huang}}, \bibinfo {author}
  {\bibfnamefont {I.}~\bibnamefont {Pletikosic}}, \bibinfo {author}
  {\bibfnamefont {A.~V.}\ \bibnamefont {Fedorov}}, \bibinfo {author}
  {\bibfnamefont {R.~D.}\ \bibnamefont {Zhong}}, \bibinfo {author}
  {\bibfnamefont {J.~A.}\ \bibnamefont {Schneeloch}}, \bibinfo {author}
  {\bibfnamefont {G.~D.}\ \bibnamefont {Gu}}, \bibinfo {author} {\bibfnamefont
  {T.}~\bibnamefont {Valla}}, \bibinfo {author} {\bibfnamefont
  {I.}~\bibnamefont {Pletikosi{\'{c}}}}, \bibinfo {author} {\bibfnamefont
  {A.~V.}\ \bibnamefont {Fedorov}}, \bibinfo {author} {\bibfnamefont {R.~D.}\
  \bibnamefont {Zhong}}, \bibinfo {author} {\bibfnamefont {J.~A.}\ \bibnamefont
  {Schneeloch}}, \bibinfo {author} {\bibfnamefont {G.~D.}\ \bibnamefont {Gu}},
  \ and\ \bibinfo {author} {\bibfnamefont {T.}~\bibnamefont {Valla}},\ }\href
  {\doibase 10.1038/nphys3648} {\bibfield  {journal} {\bibinfo  {journal} {Nat.
  Phys.}\ }\textbf {\bibinfo {volume} {12}},\ \bibinfo {pages} {550} (\bibinfo
  {year} {2016})}\BibitemShut {NoStop}%
\bibitem [{\citenamefont {Hayata}()}]{Hayata17}%
  \BibitemOpen
  \bibfield  {author} {\bibinfo {author} {\bibfnamefont {T.}~\bibnamefont
  {Hayata}},\ }\href@noop {} {\ }\Eprint {http://arxiv.org/abs/1705.09926}
  {arXiv:1705.09926} \BibitemShut {NoStop}%
\bibitem [{\citenamefont {Tse}\ and\ \citenamefont {MacDonald}(2010)}]{Tse10}%
  \BibitemOpen
  \bibfield  {author} {\bibinfo {author} {\bibfnamefont {W.-K.}\ \bibnamefont
  {Tse}}\ and\ \bibinfo {author} {\bibfnamefont {A.~H.}\ \bibnamefont
  {MacDonald}},\ }\href {\doibase 10.1103/PhysRevLett.105.057401} {\bibfield
  {journal} {\bibinfo  {journal} {Phys. Rev. Lett.}\ }\textbf {\bibinfo
  {volume} {105}},\ \bibinfo {pages} {057401} (\bibinfo {year}
  {2010})}\BibitemShut {NoStop}%
\bibitem [{\citenamefont {Maciejko}\ \emph {et~al.}(2010)\citenamefont
  {Maciejko}, \citenamefont {Qi}, \citenamefont {Drew},\ and\ \citenamefont
  {Zhang}}]{Maciejko10}%
  \BibitemOpen
  \bibfield  {author} {\bibinfo {author} {\bibfnamefont {J.}~\bibnamefont
  {Maciejko}}, \bibinfo {author} {\bibfnamefont {X.-L.}\ \bibnamefont {Qi}},
  \bibinfo {author} {\bibfnamefont {H.~D.}\ \bibnamefont {Drew}}, \ and\
  \bibinfo {author} {\bibfnamefont {S.-C.}\ \bibnamefont {Zhang}},\ }\href
  {\doibase 10.1103/PhysRevLett.105.166803} {\bibfield  {journal} {\bibinfo
  {journal} {Phys. Rev. Lett.}\ }\textbf {\bibinfo {volume} {105}},\ \bibinfo
  {pages} {166803} (\bibinfo {year} {2010})}\BibitemShut {NoStop}%
\bibitem [{\citenamefont {Li}\ \emph {et~al.}(2010)\citenamefont {Li},
  \citenamefont {Wang}, \citenamefont {Qi},\ and\ \citenamefont
  {Zhang}}]{Li10}%
  \BibitemOpen
  \bibfield  {author} {\bibinfo {author} {\bibfnamefont {R.}~\bibnamefont
  {Li}}, \bibinfo {author} {\bibfnamefont {J.}~\bibnamefont {Wang}}, \bibinfo
  {author} {\bibfnamefont {X.-L.}\ \bibnamefont {Qi}}, \ and\ \bibinfo {author}
  {\bibfnamefont {S.-C.}\ \bibnamefont {Zhang}},\ }\href {\doibase
  10.1038/nphys1534} {\bibfield  {journal} {\bibinfo  {journal} {Nat Phys}\
  }\textbf {\bibinfo {volume} {6}},\ \bibinfo {pages} {284} (\bibinfo {year}
  {2010})}\BibitemShut {NoStop}%
\bibitem [{\citenamefont {Ooguri}\ and\ \citenamefont
  {Oshikawa}(2012)}]{Ooguri12}%
  \BibitemOpen
  \bibfield  {author} {\bibinfo {author} {\bibfnamefont {H.}~\bibnamefont
  {Ooguri}}\ and\ \bibinfo {author} {\bibfnamefont {M.}~\bibnamefont
  {Oshikawa}},\ }\href {\doibase 10.1103/PhysRevLett.108.161803} {\bibfield
  {journal} {\bibinfo  {journal} {Phys. Rev. Lett.}\ }\textbf {\bibinfo
  {volume} {108}},\ \bibinfo {pages} {161803} (\bibinfo {year}
  {2012})}\BibitemShut {NoStop}%
\bibitem [{\citenamefont {Wang}\ \emph {et~al.}(2016)\citenamefont {Wang},
  \citenamefont {Lian},\ and\ \citenamefont {Zhang}}]{Wang16a}%
  \BibitemOpen
  \bibfield  {author} {\bibinfo {author} {\bibfnamefont {J.}~\bibnamefont
  {Wang}}, \bibinfo {author} {\bibfnamefont {B.}~\bibnamefont {Lian}}, \ and\
  \bibinfo {author} {\bibfnamefont {S.-C.}\ \bibnamefont {Zhang}},\ }\href
  {\doibase 10.1103/PhysRevB.93.045115} {\bibfield  {journal} {\bibinfo
  {journal} {Phys. Rev. B}\ }\textbf {\bibinfo {volume} {93}},\ \bibinfo
  {pages} {45115} (\bibinfo {year} {2016})}\BibitemShut {NoStop}%
\bibitem [{\citenamefont {Sikivie}(1983)}]{Sikivie83}%
  \BibitemOpen
  \bibfield  {author} {\bibinfo {author} {\bibfnamefont {P.}~\bibnamefont
  {Sikivie}},\ }\href {\doibase 10.1103/PhysRevLett.51.1415} {\bibfield
  {journal} {\bibinfo  {journal} {Phys. Rev. Lett.}\ }\textbf {\bibinfo
  {volume} {51}},\ \bibinfo {pages} {1415} (\bibinfo {year}
  {1983})}\BibitemShut {NoStop}%
\bibitem [{\citenamefont {Sikivie}(1985)}]{Sikivie85}%
  \BibitemOpen
  \bibfield  {author} {\bibinfo {author} {\bibfnamefont {P.}~\bibnamefont
  {Sikivie}},\ }\href {\doibase 10.1103/PhysRevD.32.2988} {\bibfield  {journal}
  {\bibinfo  {journal} {Phys. Rev. D}\ }\textbf {\bibinfo {volume} {32}},\
  \bibinfo {pages} {2988} (\bibinfo {year} {1985})}\BibitemShut {NoStop}%
\bibitem [{\citenamefont {Wang}\ \emph {et~al.}(2011)\citenamefont {Wang},
  \citenamefont {Li}, \citenamefont {Zhang},\ and\ \citenamefont
  {Qi}}]{Wang11}%
  \BibitemOpen
  \bibfield  {author} {\bibinfo {author} {\bibfnamefont {J.}~\bibnamefont
  {Wang}}, \bibinfo {author} {\bibfnamefont {R.}~\bibnamefont {Li}}, \bibinfo
  {author} {\bibfnamefont {S.-C.}\ \bibnamefont {Zhang}}, \ and\ \bibinfo
  {author} {\bibfnamefont {X.-L.}\ \bibnamefont {Qi}},\ }\href {\doibase
  10.1103/PhysRevLett.106.126403} {\bibfield  {journal} {\bibinfo  {journal}
  {Phys. Rev. Lett.}\ }\textbf {\bibinfo {volume} {106}},\ \bibinfo {pages}
  {126403} (\bibinfo {year} {2011})}\BibitemShut {NoStop}%
\bibitem [{\citenamefont {Lee}\ \emph {et~al.}(2015)\citenamefont {Lee},
  \citenamefont {Park}, \citenamefont {Ihm},\ and\ \citenamefont
  {Son}}]{Lee15}%
  \BibitemOpen
  \bibfield  {author} {\bibinfo {author} {\bibfnamefont {Y.-L.}\ \bibnamefont
  {Lee}}, \bibinfo {author} {\bibfnamefont {H.~C.}\ \bibnamefont {Park}},
  \bibinfo {author} {\bibfnamefont {J.}~\bibnamefont {Ihm}}, \ and\ \bibinfo
  {author} {\bibfnamefont {Y.-W.}\ \bibnamefont {Son}},\ }\href {\doibase
  10.1073/pnas.1515664112} {\bibfield  {journal} {\bibinfo  {journal}
  {Proceedings of the National Academy of Sciences}\ }\textbf {\bibinfo
  {volume} {112}},\ \bibinfo {pages} {11514} (\bibinfo {year}
  {2015})}\BibitemShut {NoStop}%
\bibitem [{\citenamefont {{Dzyaloshinskii, I}}(1960)}]{DzyaloshinskiiI60}%
  \BibitemOpen
  \bibfield  {author} {\bibinfo {author} {\bibfnamefont {E.}~\bibnamefont
  {{Dzyaloshinskii, I}}},\ }\href@noop {} {\bibfield  {journal} {\bibinfo
  {journal} {Sov. Phys. JETP}\ }\textbf {\bibinfo {volume} {10}},\ \bibinfo
  {pages} {628} (\bibinfo {year} {1960})}\BibitemShut {NoStop}%
\bibitem [{\citenamefont {Astrov}(1960)}]{Astrov60}%
  \BibitemOpen
  \bibfield  {author} {\bibinfo {author} {\bibfnamefont {D.~N.}\ \bibnamefont
  {Astrov}},\ }\href@noop {} {\bibfield  {journal} {\bibinfo  {journal} {Sov.
  Phys. JETP}\ }\textbf {\bibinfo {volume} {11}},\ \bibinfo {pages} {708}
  (\bibinfo {year} {1960})}\BibitemShut {NoStop}%
\bibitem [{\citenamefont {Fiebig}(2005)}]{Fiebig05}%
  \BibitemOpen
  \bibfield  {author} {\bibinfo {author} {\bibfnamefont {M.}~\bibnamefont
  {Fiebig}},\ }\href {http://stacks.iop.org/0022-3727/38/i=8/a=R01} {\bibfield
  {journal} {\bibinfo  {journal} {Journal of Physics D: Applied Physics}\
  }\textbf {\bibinfo {volume} {38}},\ \bibinfo {pages} {R123} (\bibinfo {year}
  {2005})}\BibitemShut {NoStop}%
\bibitem [{\citenamefont {Garate}\ and\ \citenamefont
  {Franz}(2010)}]{Garate10}%
  \BibitemOpen
  \bibfield  {author} {\bibinfo {author} {\bibfnamefont {I.}~\bibnamefont
  {Garate}}\ and\ \bibinfo {author} {\bibfnamefont {M.}~\bibnamefont {Franz}},\
  }\href {\doibase 10.1103/PhysRevLett.104.146802} {\bibfield  {journal}
  {\bibinfo  {journal} {Phys. Rev. Lett.}\ }\textbf {\bibinfo {volume} {104}},\
  \bibinfo {pages} {146802} (\bibinfo {year} {2010})}\BibitemShut {NoStop}%
\bibitem [{\citenamefont {Yokoyama}\ \emph {et~al.}(2010)\citenamefont
  {Yokoyama}, \citenamefont {Zang},\ and\ \citenamefont
  {Nagaosa}}]{Yokoyama10a}%
  \BibitemOpen
  \bibfield  {author} {\bibinfo {author} {\bibfnamefont {T.}~\bibnamefont
  {Yokoyama}}, \bibinfo {author} {\bibfnamefont {J.}~\bibnamefont {Zang}}, \
  and\ \bibinfo {author} {\bibfnamefont {N.}~\bibnamefont {Nagaosa}},\ }\href
  {\doibase 10.1103/PhysRevB.81.241410} {\bibfield  {journal} {\bibinfo
  {journal} {Phys. Rev. B}\ }\textbf {\bibinfo {volume} {81}},\ \bibinfo
  {pages} {241410} (\bibinfo {year} {2010})}\BibitemShut {NoStop}%
\bibitem [{\citenamefont {Sakai}\ and\ \citenamefont {Kohno}(2014)}]{Sakai14}%
  \BibitemOpen
  \bibfield  {author} {\bibinfo {author} {\bibfnamefont {A.}~\bibnamefont
  {Sakai}}\ and\ \bibinfo {author} {\bibfnamefont {H.}~\bibnamefont {Kohno}},\
  }\href {\doibase 10.1103/PhysRevB.89.165307} {\bibfield  {journal} {\bibinfo
  {journal} {Phys. Rev. B}\ }\textbf {\bibinfo {volume} {89}},\ \bibinfo
  {pages} {165307} (\bibinfo {year} {2014})}\BibitemShut {NoStop}%
\end{thebibliography}%

\end{document}